\documentstyle[12pt]{article}
\def\beq{\begin{eqnarray}}
\def\endeq{\end{eqnarray}}
\def\nn{\nonumber \\}
\def\dalembert{{\sqcap\!\!\!\!\sqcup}}
\def\eps{\epsilon}
\def\pd{\partial}
\def\e{{\rm e}}

\tiny\normalsize
\begin{document}
\centerline{\bf CAN A GRAVITATIONAL WAVE AND A MAGNETIC}
\centerline{\bf MONOPOLE COEXIST?}
\vskip .5cm
\centerline{Re: MPLA/358/97}
\vskip 5mm
\centerline{\footnotesize OSAMU ABE\footnote{e-mail address: 
osamu@asa.hokkyodai.ac.jp}
and
OSAMU TABATA\footnote{e-mail address: 
x3920@atson.asa.hokkyodai.ac.jp}}
\centerline{\it Physics Laboratory, Asahikawa Campus}
\centerline{\it Hokkaido University of Education, 9 Hokumoncho}
\centerline{\it Asahikawa 070, Japan}
\begin{abstract}
We investigate the behavior of small perturbations around the 
Kaluza-Klein monopole in the five dimensional space-time. We find that
the even parity gravitational wave does not propagate in the five 
dimensional space-time with Kaluza-Klein monopole provided that the 
gravitational wave is constant in the fifth direction.
We conclude that a gravitational wave and a U(1) magnetic monopole 
do not coexist in five dimensional Kaluza-Klein spacetime.
\end{abstract}
\vskip 1cm
{\noindent\normalsize PACS numbers: 03.40.Kf, 04.30.+x, 11.10.Kk}
%%%%%%%%%%%%%%%%%%%%%%%%%%%  Introduction  %%%%%%%%%%%%%%%%%%%%%%%%%
\newpage
%\section{Introduction}
Dirac \cite{dirac} proposed the theory of the magnetic monopole in 
1931. Dirac's conjecture of monopole is attractive because it can 
explain the 
quantization of electric charge of elementary particles. In 1974, 
't~Hooft \cite{thooft} and Polyakov \cite{polyakov} independently 
found that a certain class of non-abelian gauge theories not only 
allow but also demand the existence of superheavy magnetic monopoles.
Then most of all grand unified theories demand 't~Hooft-Polyakov 
monopoles with mass comparable to the unification mass which is 
typically $10^{15-17}{\rm GeV}$. 

Thus, we may expect that superheavy magnetic monopoles could have been 
generated just after the big bang, the birth of our universe and 
that there must be monopoles in the cosmic rays. 
Many researches have been being made for finding magnetic 
monopoles \cite{giacomelli}, but no good candidate has been found so 
far except for the observation which was made in 1982 \cite{cabrera}.

The simpler grand unified theories requires too many magnetic 
monopoles, and contradict above experimental results.
Guth \cite{guth} gave a nice mechanism, which is called as the 
inflationary universe scenario, that might suppress 't~Hooft-%
Polyakov monopole production.

In the present paper, we would like to discuss a new mechanism 
that denies the existence of Dirac's U(1) magnetic monopoles in 
connection with a gravitational wave.
Taylor \cite{taylor} has shown an indirect verification of the 
existence of gravitational radiation. Though there is no direct 
evidence, the existence of the gravitational wave is beyond doubt.

   Our discussion is based on the Kaluza-Klein theory \cite{k-k}. 
Originally, Kaluza and Klein suggested the possibility that 
gravitation and electromagnetism could be unified in a theory in 
the framework of five-dimensional space-time.
DeWitt \cite{dewitt} and others \cite{others} presented its 
generalization to non-abelian 
gauge group in 4+d dimensional space-time.
Gross and Perry \cite{gross} and Sorkin \cite{sorkin}
found the existence of the regular, static and topologically 
stable magnetic monopoles solution of the five dimensional 
Einstein equation. The existence of the magnetic monopole strongly 
indicates that the fundamental interactions are unified. Sundaresan 
and Tanaka  \cite{tanaka} discussed the stability of the of the K-K 
monopole against the even parity perturbation with $l =m=0$ in the 
positive definite metric, i.e. Euclidean space-time. In Ref. 
\cite{abe1}, 
one of present authors discussed the existence of gravitational wave 
in indefinite metric without the magnetic monopole. One of present 
authors studied \cite{abe2} the odd parity perturbations around the 
magnetic monopole solution, and concluded that  there is no parity 
odd gravitational wave.

In this paper, we will investigate the 
propagation of the parity even
perturbations around the K-K monopole solution in the indefinite 
metric, i.e. Minkowskian space-time.

   We start with a theory of gravitation described by the 
Einstein-Hilbert action in five dimensions
\beq
S={-1\over 16\pi G_5}\int d^5x\sqrt{|g_5|}R_5,
\label{einstein}
\endeq
where $G_5$ is the five dimensional gravitational constant, $g_5$ is 
the determinant of five dimensional metric $g_{AB}$ and $R_5$ denotes 
the five dimensional curvature scalar of the space-time. 

   The magnetic monopole solution is the solution of the 
Einstein equation in the empty space-time
\beq
R_{AB} = 0.
\endeq
We take the metric $g_{AB}$ as \cite{gross}
\beq
ds^2=g_{AB}dx^Adx^B,\; x^A=(t,r,\theta ,\varphi ,x^5),
\endeq
where
\beq
g_{AB}=\left(\matrix{
     1& 0&          0&                        0&           0\cr
     0& -V^{-1}&    0&                        0&           0\cr
     0& 0& -r^2V^{-1}&                        0&           0\cr
     0& 0&          0& -{r^2s^2+A^2V^2 \over V}&         -AV\cr
     0& 0&          0&                      -AV&          -V\cr
     }\right)
\label{metric}
\endeq
Here,
\beq
V^{-1}(r) = 1 + {4M\over r},\; s = \sin\theta .
\endeq
The gauge field, $A_\mu$, is that of monopole at the origin
\beq
A_\mu (x) = \Bigl(A_t, A_r, A_\theta , A_\varphi=A\Bigr)
=\Bigl(0,0,0,4M\alpha (\theta )\Bigr),
\endeq
where
\beq
\alpha(\theta ) =\cases{
1-\cos\theta&in $R_a:\{0\le r,0\le\theta<{\pi\over 2}+\delta ,
0\le\varphi <2\pi\}$\cr
-1-\cos\theta&in $R_b:\{0\le r,{\pi\over 2}-\delta <\theta\le\pi,
0\le\varphi <2\pi\}$\cr},
\endeq
and $A_\mu$ satisfies
\beq
{\bf B} ={\bf \nabla}\times {\bf A} = {4M{\bf r}\over r^3}.
\endeq
In order to avoid the so-called NUT singularity, $M$ should 
satisfy \cite{gross}
\beq
M = {\sqrt{\pi G}\over 2e},
\endeq
where $G$ is the four dimensional gravitational constant and $e$ 
denotes the unit electric charge. Then the magnetic 
charge $g$ of the monopole is given by \cite{gross}
\beq
     g = {4M\over \sqrt{16\pi G}} ={1\over 2e}.
\endeq
Thus the monopole has one unit of Dirac charge.

   Now, we consider small perturbations around the monopole 
solution. Then the total metric $g^T_{AB}$ is given by
\beq
 g^T_{AB} = g_{AB} + h_{AB}.
\endeq
According to the approach of Regge and Wheeler \cite{regge}, 
we can divide small perturbation into even and odd parity:
\beq
& &h_{AB}^{even}(t,r,\theta ,\varphi , x^5)=\nn 
& &  \left(\matrix{
  H_0(r)& H_1(r)& H_3(r)\nabla_2 & H_3(r)\nabla_3 & H_{0p}(r)\cr
  Sym   & H_2(r)& H_4(r)\nabla_2 & H_4(r)\nabla_3 & H_{1p}(r)\cr
  Sym   & Sym   & r^2\gamma_{22}K(r)+L(r)\nabla_2\nabla_2  
                    & L(r)\nabla_2\nabla_3 & H_5(r)\nabla_2\cr
  Sym   & Sym   & Sym & r^2\gamma_{33}K(r)+L(r)\nabla_3\nabla_3
                                             &  H_5(r)\nabla_3 \cr
  Sym   & Sym   & Sym                 & Sym         & H_{2p}(r) \cr
                                                 }\right)\nn 
& &  \times   Y_{qlm}(\theta ,\varphi )\e^{-i\omega t}\e^{in^5x^5/R}
\label{even}
\endeq
and
\beq
& &h_{AB}^{odd}(t,r,\theta ,\varphi ,x^5)=\nn
& & \left(\matrix{
  0&   0& {\eps_2}^3h_0(r)\nabla_3& {\eps_3}^2h_0(r)\nabla_2 & 0\cr
  Sym& 0& {\eps_2}^3h_1(r)\nabla_3& {\eps_3}^2h_1(r)\nabla_2 & 0\cr
  Sym& Sym& 2{\eps_2}^3h_2(r)\nabla_3\nabla_2& 
      h_2(r)[{\eps_2}^3\nabla_3\nabla_3+{\eps_3}^2\nabla_2\nabla_2]&
	                                 {\eps_2}^3h_3(r)\nabla_3\cr
  Sym& Sym&    Sym&    2{\eps_3}^2h_2(r)\nabla_2\nabla_3&
                                         {\eps_3}^2h_3(r)\nabla_2\cr
  Sym& Sym&    Sym&    Sym&                                     0\cr
                                                   }\right)\nn 
& & \times Y_{qlm}(\theta ,\varphi )\e^{-i\omega t}\e^{in^5x^5/R}.
\label{odd}
\endeq
Here,  $R$ is the radius of the circle in the fifth dimension and we 
have used
\beq
\gamma_{AB} = {g_{AB}\over r^2},
\endeq
\beq
{\eps_2}^3 \equiv -g^{3A}\sqrt{|g_5|}e_{012A5} = -{1\over s},
\endeq
and
\beq
{\eps_3}^2 \equiv -g^{2A}\sqrt{|g_5|}e_{013A5} = s,
\endeq
where $e_{ABCDE}$'s are totally antisymmetric Levi-Chivita symbols in 
five dimensions. In Eqs.~(\ref{even}) and (\ref{odd}), $Y_{qlm}$ 
denotes the monopole harmonics \cite{wu}
which is introduced so as to avoid the singularity at $\theta =\pi$ 
and which satisfies following eigen value equations \cite{dray}.
\beq
{\bf L}^2Y_{qlm}
&\equiv &\Bigl\{-{1\over s}\pd_\theta s\pd_\theta 
-{1\over s^2}\pd_\varphi^2-{2q\alpha (\theta )L_z\over s^2}
   \Bigr\}Y_{qlm}
=l (l +1)Y_{qlm},\nn 
 L_zY_{qlm}&\equiv &\Bigl\{ -i\pd_\varphi -q\alpha (\theta )-q\cos
\theta\Bigr\}
Y_{qlm}=mY_{qlm}.
\endeq
The monopole harmonics $Y_{qlm}$ coincides with the usual spherical 
harmonics $Y_{l m}$ when $q=0$. The number $q$ is defined by $q=eg$ 
and which equals to ${1\over 2}$ in the case of K-K monopole.

   We will consider small perturbations with $n^5=0$ that is we 
consider only perturbations which are constant in the fifth direction.
We impose the gauge conditions or transversality conditions on the 
small perturbations as follows
\beq
\nabla_Bh^{AB} = 0,
\endeq
where $h^{AB}=-g^{AC}g^{BD}h_{CD}$.
   We derive equations for small perturbation $h$ as follows. The 
Ricci tensor will be $R_{AB}$ if it is calculated from $g_{AB}$ 
and $R_{AB}+\delta R_{AB}$ if it is calculated from $g_{AB}+h_{AB}$. 
We can get second order equations on the perturbation $h$ if we 
impose the condition $\delta R_{AB}=0$. 
This condition implies that the perturbed space is also empty space. 
The explicit form of the equations are
\beq
\delta R_{AB}={{{R_A}^C}_B}{^D}h_{CD}+{\dalembert\over 2}h_{AB} = 0.
\endeq

Now, we will investigate whether the even parity 
perturbations can propagate in the five dimensional space-time with 
K-K monopole provided that the perturbations are constant in the 
fifth direction.
First, we consider $(0,0)$ component of $\delta R$. We have
\beq
0&=&2\delta R_{00}\nn
&=&-\e^{i\omega t}V\; Y_{qlm}
\Bigl( H_0''+{2\over r}H_0'-{l (l +1)\over r^2}H_0
+{\omega^2\over V}H_0-{2qm\alpha \over s^2r^2}H_0\Bigr).
\label{R00}
\endeq
We can separate $\alpha$ dependent part from $\alpha$ independent 
parts using the identities
\beq
\int_0^\pi d\theta \cos\theta =0,\quad
\int_0^\pi d\theta{\alpha (\theta )\cos\theta\over \sin^2\theta}
    =\pi -2.
\endeq
Thus, we can conclude $H_0\equiv 0$ from Eq.~(\ref{R00}).

Next, we consider $(0,5)$ component of the Ricci tensor
\beq
0&=&2\delta R_{05}=\nn
& &-V\e^{-i\omega t}
\Bigl\{
Y_{qlm}\Bigl[H_{0p}''+{1+V\over r}H_{0p}'-{l (l +1)\over r^2}H_{0p}
+{\omega^2\over V}H_{0p}\Bigr]\nn
& &-{2qm\alpha \over s^2r^2} Y_{qlm}H_{0p}
+{(V-1)^2\alpha \over sr^2}
\pd_\theta Y_{qlm}H_{0p}\Bigr\}.
\label{R05}
\endeq
If the functions $Y_{qlm}$, $Y_{qlm}\alpha/s^2$, and 
$\pd_\theta Y_{qlm}/s$ 
are linearly independent, i.e. the Wronskian is not vanishing, each 
coefficient must vanish. We can prove that above functions are 
linearly independent. Thus we can conclude that $H_{0p}\equiv 0$ from 
Eq.~(\ref{R05}). 

Further, we consider $(1,5)$ component. That is
\beq
0&=&2\delta R_{15}=\nn
&-&V\e^{-i\omega t}\Bigl\{Y_{qlm}\Bigl[H_{1p}''+{2\over r}H_{1p}'
-{l(l+1)-(V^2-2V-1)\over r^2}H_{1p}+{\omega^2\over V}H_{1p}\nn
& &+{l(l+1)(V+1)\over r^3}H_5\Bigr]\nn
&+&{2qm\alpha\over s^2 r^2}Y_{qlm}\Bigl[-H_{1p}+{(V+1)\over r}H_5
\Bigr]
-{4M\alpha V(V-1)\over sr^3}\pd_\theta Y_{qlm}H_{1p}\nn
&+&{4M\alpha(V+1)\over s^2r^3}\pd_\varphi Y_{qlm}H_{2p}
\Bigr\}
\endeq
Analogous argument for Eq.~(\ref{R05}) is applicable even in this 
case. That is
$\pd_\varphi Y_{qlm}$ is linearly 
independent in addition to three functions those are appeared in 
Eq.~(\ref{R05}). Thus we have
$H_{1p}\equiv 0$, $H_{2p}\equiv 0$ and then $H_5\equiv 0$.

The transversality condition for 0-th component becomes
\beq
0=-\nabla_Ah^{A0}={V\e^{-i\omega t}\over r^2}\Bigl\{
Y_{qlm}\left[ l(l+1)H_3-(r^2H_1)'\right]
+{2qm\alpha\over s^2}Y_{qlm}H_3\Bigr\}.
\endeq
From which we obtain $H_3\equiv 0$ and $H_1 = {\rm const.}/r^2$.

We can solve $\delta R_{02}=0$ for $H_1$.
\beq
0=2\delta R_{02}=-{V(V+1)\e^{-i\omega t} \over r}\pd_\theta 
Y_{qlm}H_1,
\endeq
and have $H_1\equiv 0$.

Next, we consider the traceless conditon. Because $H_0=0$, $H_5=0$, 
and 
$H_{2p}=0$, the traceless condition becomes
\beq 
0&=&{\rm Tr}h^{even}=\nn
&-&{\e^{-i\omega t}\over s^2 r^3 }\Bigl\{
s^2rY_{qlm}\bigl[r^2VH_2-2r^2K-l(l+1)VL\bigr]
-16M^2rV^2\alpha^2 Y_{qlm}K\nn
&-&2rVqm\alpha Y_{qlm}L
+4V^2(V-1)Ms\alpha \pd_\theta Y_{qlm}L
\Bigr\}.
\label{trh}
\endeq
The same argument as is used for Eq.~(\ref{R05}) is applicable in 
this case. Thus, we obtain $L\equiv 0$, $K\equiv 0$, and 
$H_2\equiv 0$.

Finally, we consider the transversality condition for 1-st component.
\beq
0=\nabla_Ah^{A1}={V^2\e^{-i\omega t}\over r^2}\left[
l(l+1)+{2qm\alpha\over s^2}\right]Y_{qlm}H_4.
\endeq
Thus, we have $H_4\equiv 0$. 
We have proven that the small 
perturbations which are appeared in Eq.~(\ref{even}) are all vanishing. 

%\section{Summary and conclusion}
In this paper, we have studied the properties of the small 
perturbations around the five dimensional Kaluza-Klein metric with 
magnetic monopole. 
We found that there is no even parity perturbation provided that the 
perturbation is independent of the fifth coordinate. 
Combining our result with the result in Ref.~\cite{abe2}, we can 
conclude that a gravitational wave can not coexist with 
Kaluza-Klein magnetic monopole in the five dimensional space-time.
If we believe the existence of the gravitational wave, our result
implies the absence of cllasical magnetic monopole.
\vfill\eject\noindent
{\Large\bf Acknowledgements}
\vskip.5cm
\noindent
One of present authors(O.A.) acknowledges useful discussions with 
Prof. K. Tanaka and Prof. M. K. Sundaresan. He also would like to 
thank the Ohio State University, Department of Physics for their 
hospitality, where this work was started. 

\end{document}